\documentclass[english,twocolumn,nofootinbib]{revtex4-1}
\usepackage{amsmath,amssymb,amsfonts}
\usepackage{comment}
\usepackage{lipsum}
\usepackage{epsfig,verbatim,graphics}
\usepackage{subfigure,slashed}

\usepackage{hyperref}
\usepackage[utf8]{inputenc}
\usepackage[english]{babel}
\usepackage{geometry}
\begin{document}
	\hfill Nikhef 2016-055
		\title{\Large {Renormalization Group independence of Cosmological Attractors}}
		\author{Jacopo Fumagalli}\email{jacopof@nikhef.nl}
		\address{Nikhef, Science Park 105, 1098 XG Amsterdam, The Netherlands}

	\begin{abstract}

		The large class of inflationary models known as $\alpha$- and
		$\xi$-attractors gives identical cosmological predictions at tree level (at
		leading order in inverse power of the number of efolds). Working
		with the renormalization group improved action, we show that these
		predictions are robust under quantum corrections. This means that for all the models considered the inflationary parameters $(n_s,r)$ are (nearly) independent on the Renormalization Group flow. The result
		follows once the field dependence of the renormalization scale,
		fixed by demanding the leading log correction to vanish, satisfies a
		quite generic condition.  In Higgs inflation (which is a particular $\xi$-attractor) this is indeed
		the case; in the more general attractor models this is still
	    ensured by the renormalizability of the theory in the effective
		field theory sense.

	\end{abstract}

\maketitle

\section{Introduction}

Successful inflationary models should satisfy some basic
requirements. They have to be consistent within the theory in which
they are formulated (QFT and GR). Moreover, for the models to be
predictive, the predictions should depend on a number of parameters
smaller than the number of predictions themselves.

A large class of inflationary models, the so-called Cosmological
Attractors, gives the same classical predictions for the inflationary
observables\footnote{More precisely, for $\alpha\rightarrow1$ and
  $\xi\rightarrow\infty$ in the $\alpha$- and $\xi$-attractors
  respectively.} \cite{Galante:2014ifa}. At leading order in the $1/N$
expansion, with $N$ the number of efolds, the tree-level spectral
index and tensor-to-scalar ratio are given by $n_{s}=1-2/N+O(N^{-2})$
and $r=12/N^{2}+O(N^{-3})$. A natural question to ask then is if these
predictions are robust in the full quantum theory. Are the attractor
models consistent and predictive at the quantum level?  In this letter
we will consider the effect of perturbative corrections due to the
renormalization group (RG) flow.

In single field inflation, the quantum corrected dynamics of the
inflaton is given by an effective action of the form 
\begin{equation}
\frac{\mathcal{L}_{{\rm eff}}}{\sqrt{-g}}=-\tfrac{1}{2}Z(\phi)K_{\phi}(\phi)(\partial\phi)^{2}-V(\phi)+{\rm ...}
\end{equation}
with $Z(\phi)$ the (non-trivial) renormalization wavefunction,
$K_{\phi}$ the metric in field space in presence of a non-canonical
kinetic term, $V$ the full quantum potential and the dots stand for
higher derivative terms that can be safely neglected in the slow roll
approximation.  With the gravity sector in the standard form, i.e. for
the action in the Einstein frame, the slow roll parameters as well as
the inflationary indexes $n_{s}$ and $r$ are given in terms of
derivatives of the effective potential with respect to the canonical
inflaton field.  Using standard renormalization group techniques it is
possible to rewrite the effective action in a form suitable for the
inflationary analysis, which takes into account the leading log
expansion of the quantum potential.

We are interested in the possibility that quantum corrections
enter at first order in the $1/N$-expansion. The observables can then
be written in the general form
\begin{equation}
n_{s}\simeq1-\frac{2}{N}f_{n}(\beta_{\lambda_{i}},\lambda_{j})_{\star}\,,
\quad r\simeq\frac{12}{N^{2}}f_{r}(\beta_{\lambda_{i}},\lambda_{j})_{\star},
\end{equation}
evaluated at the time the pivot scale
($k_{\star}=0.002\,{\rm Mpc^{-1}}$) leaves the horizon (denoted by a
subscript $\star$).  Here $f_{n}$ and $f_{r}$ are two generic
functions of the beta functions $\beta_i$ and the couplings of the
model $\lambda_i$. If the inflationary parameters have such a
dependence,\footnote{Note that even though $\beta_{\lambda_{i}}$ might
  be small in general, this is not necessarily true for combinations
  of $(\beta_{\lambda_{i}},\lambda_{j})$,
  e.g. $\beta_{\lambda_i}/\lambda_i$.} this would imply that the knowledge
of the details of the renormalization group (RG) flow during inflation
are needed (to find out the expressions for $\beta_{\lambda_{i}}$) to
draw conclusions on the model.  Even more, to ever connect the low and
high energy regimes of the model, one would need to know the details
of the RG flow through the entire energy domain.

The RG dependence of the observables can be both a curse and a
blessing. A blessing because it can lift the degeneracy between the
different attractor models.  Moreover, including loop corrections to
the inflaton action could in principle shed light on the UV dependence
of the inflationary parameters. On the other hand, if a model depends
strongly on unknown UV corrections it will lose completely any
predictive power. 
In this letter, generalizing the idea of our previous work \cite{Fumagalli:2016lls}, we show that for the large class of
inflationary attractor models, the $\alpha$- and $\xi$-attractors
\cite{Ferrara:2013rsa,Kallosh:2013yoa,Kallosh:2013tua,Giudice:2014toa,Csaki:2014bua},
$n_{s}$ and $r$ are nearly independent on the RG flow. This also
implies that any kind of UV physics whose effect enters only via the
RG flow (i.e., that does not affect the inflationary potential already at tree
level) will in general have no effect on the predictions for these
models.

\subsection{RG improving and renormalization scale\label{sub:RG-improving-and}}
Let us briefly review some standard features of the effective action
and the RG flow that we will use in the following. The quantum
potential for a scalar field $\phi$ depends in general on powers of
logarithms of the form $\ln\left({M_{i}^{2}(\phi)}/{\mu^{2}}\right)$
where $M_{i}^{2}(\phi)$ are the field-dependent masses of the
particles running in the loops and $\mu$ the renormalization
scale. The logarithms appear only up to $L$-th power at the $L$ loop
order. A well known result in quantum field theory
\cite{Bando:1992np,Bando:1992wy} tells us that in each region of the
field space it is possible to define an effective field theory (EFT)
where only one logarithm remains relevant in the full effective
potential. All the other mass scales decouple and their net effect
will be a shift in the definition of the parameters of the EFT. Thus,
schematically, each loop contribution will have the following form
\begin{equation}
V^{(L)}=\hbar^{-1}(v_{0}^{(L)}s^{L}+\hbar v_{1}^{(L)}s^{L-1}..+\hbar^{L}v_{L}^{(L)}),
\end{equation}
where 
\begin{equation}
s=\hbar\ln\left(\frac{M^{2}(\phi)}{\mu^{2}}\right)
\end{equation}
is the only relevant log in the EFT. Here $v_{i}^{(j)}$ are functions
of the field and all the other couplings/mass parameters $(\lambda_{i})$,
and $\hbar$ is the loop counting parameter. The full potential can
be written in general as \cite{Kastening:1991gv} 
\begin{equation}\begin{split}
V&=M^{4}(\phi)\sum_{i=0}^{\infty}\hbar^{i-1}\left(\sum_{L=i}^{\infty}v_{i}^{(L)}s^{L-i}\right)\\[1mm]
&\equiv M^{4}(\phi)\sum_{i=0}^{\infty}\hbar^{i-1}f_{i},
\label{eq:log exp}
\end{split}\end{equation}
where $f_{i}$ is the $i$-th to leading log term. We label the
potential truncated at $L$-loop order with
$V_L=V^{(0)}+...+V^{(L)}$. $V$ satisfies the renormalization group
equation (RGE) \cite{Callan:1970yg,Symanzik:1970rt}:
\begin{equation}\label{callan}
\mathcal{D}V\equiv\left(\mu\frac{\partial}{\partial\mu}+\beta_{\lambda_{i}}\frac{\partial}{\partial\lambda_{i}}-\gamma\phi\frac{\partial}{\partial\phi}\right)V=0,
\end{equation}
where $\gamma$ is the anomalous dimension of the scalar field. This
allows us to rewrite it as a formal solution in the following standard
way (see for example \cite{Ford:1992mv})
\begin{equation}
V(\phi,\lambda_{i},\mu)=V(\bar{\phi}(t),\bar{\lambda}_{i}(t),\bar{\mu}(t))\label{eq:solution RGE}
\end{equation}
where 
\begin{equation}\begin{split}
\frac{d\bar{\phi}(t)}{dt}&=-\gamma(\bar{\lambda}_{i}(t))\bar{\phi}(t)\,,\quad\frac{d\bar{\lambda}_{i}(t)}{dt}=\beta_{i}(\bar{\lambda}_{j}(t)),\\[1mm]
\bar{\mu}(t)&=\mu e^{t},
\end{split}\end{equation}
and with the initial conditions that the barred quantities reduce to the
unbarred ones at $t=0$. In a standard quantum field theory the renormalizability
ensures that in each EFT the RG operators $\mathcal{D}$ are the same.
Indeed, we are actually solving the same equation by simply using
different set of parameters. The matching between the solutions is
provided by the equations that relate the parameters of two adjacent
EFTs evaluated at the renormalization point $\mu$ around the threshold
\cite{Bando:1992wy}.

Given a generic inflationary model renormalizable in the EFT sense in
the inflationary regime,\footnote{In each field region it is possible
  to define a small parameter; there should be a finite number of
  counter terms at every order in the expansion in this small
  parameter.} the operators $\mathcal{D}$ are not necessarily the same
in each EFT (defined at different energy scales). To patch together
the EFTs we would then need some threshold corrections, i.e. some
extra UV physics \cite{Fumagalli:2016lls,Hertzberg2,cliffnew}.  This
is for example the case in Higgs inflation
\cite{bezrukov1,Bezrukov:2013fka} where the beta functions are
different in the low and middle/large regime
\cite{bezrukov_loop,wilczek,barvinsky,barvinsky2,barvinsky3,damien2}.
The UV physics can be parameterized by a tower of higher order
operators, which have a net effect on the boundaries of the EFTs and
as such provide the necessary threshold corrections \cite{burgess2}.
In this way the UV physics can enter the predictions through the RG
flow, that is different Wilson coefficients for the higher order
operators could result in different $n_{s}$ and $r$. As we will show
in this paper, it turns out that for a large class of inflationary
models we do not really need to know the details of the RG flow to
derive the inflationary parameters.

When we formally solve the RG equation in the inflationary regime,
only one log remains relevant. Equation \eqref{eq:solution RGE} tells
us that the effective potential is determined once its functional form
is known for a certain value of $t$. The standard procedure to derive useful information from
\eqref{eq:solution RGE}, is to choose $t$ in such a way
that
\begin{equation}
\bar{s}(\tilde{t})=\ln\frac{\bar{M}^{2}(\tilde{t})}{\bar{\mu}^{2}(\tilde{t})}=0\,\implies\,\tilde{t}(\phi,\lambda_{i}(\tilde{t}))\label{eq:sbar}
\end{equation}
where we omit from now on the bar over the running couplings. In this
way
$V=V(t)|_{\bar{s}=0}=V_{L}(t)|_{\bar{s}=0}+O(\hbar^{L})$.\footnote{It
  might seem obvious that $V$ and $V_{L}$ differ by $O(\hbar^{L})$
  terms. However, what we mean by that are order $\hbar^{L}$ terms in
  the leading log series expansion \eqref{eq:log exp}.} This means
that the knowledge of the $L$ loop potential (and the function
$\tilde{t}$) provides an exact RG improved potential up to order $L$
in this leading log expansion. Therefore, depending on the order we
want to work at, the potential used in our computations is given by
\begin{equation}
V_{L}|_{\bar{s}=0}\equiv V_{L}(\tilde{t}),
\end{equation}
and the RGE coefficients functions $\beta,\gamma$ at $(L+1)$-loop
order \cite{Bando:1992np,Bando:1992wy}.
We will consider the leading corrections, that is we set $L=0$ and use
the 1-loop $\beta$-functions. Even if our results will not depend on the loop order of the $\beta$-functions considered, this does not imply that it holds automatically beyond the leading order. In section \ref{sec:higher} we
comment on the generalization of our results to higher orders.

Let us make an important remark here.  In the following (we used this
already in \eqref{eq:sbar}) we will consider $\phi$ instead of its
barred and $t$ dependent version $\bar{\phi}(t)$ in the RG improved
potential (and $\rho$ instead of $\bar{\rho}(t)$ in the next
sections). We are allowed to do this for the following reason (see
\cite{Espinosa:2015qea} and our appendix A in \cite{Fumagalli:2016lls}
for more details). Consider the improved renormalization wavefunction
as absorbed in the field redefinition, i.e.
$Z_{{\rm eff}}(t)(\partial\phi)^{2}=(\partial\phi_{{\rm can}})^{2}$.
Then we have, at leading order
\begin{equation}
\bar{\phi}=e^{-\int\gamma dt'}\phi=e^{-\int\gamma dt'}Z_{{\rm
    eff}}^{-1/2}\phi_{{\rm can}}\approx\phi_{{\rm can}},
\label{fn:Remark unbarred field}
\end{equation}
where we simply omit the subscript ``can''.

\subsection{Key idea}

The key point is that since only one log remains relevant during
inflation there is (up to some irrelevant numerical factors) a unique
choice for the function $\tilde{t}$, i.e. the one implicitly defined
by eq. \eqref{eq:sbar}. In order to compute the inflationary
parameters $n_s$ and $r$ we take derivatives of the effective
potential with respect to the scalar field. These will be a function
of derivatives of $\tilde{t}$ as well as of the couplings and the
$\beta$-functions. Thus the predictions can in principle depend on the
RG flow during inflation (through the value of the beta functions in
this regime) and on the full RG flow (through the value of the running
couplings at the horizon). Expanding the equations in powers of the small parameter $\rho$
defining the inflationary regime it can be shown analytically, without
having to solve explicitly the RG equations, that for the inflationary
Cosmological Attractors models, neither of the two contributions
influence the inflationary predictions at first order in inverse power
of the number of efolds.\footnote{The scalar power spectrum constrains
  one combination of couplings in the theory, which can always be
  satisfied fixing the free parameter of the model. For different RG
  evolutions, the individual couplings may have different values at
  horizon exit, but as long as the combination is kept fixed by
  adjusting the free parameter, this has no direct observable
  consequences.}


\section{Inflationary parameters}

\subsection{General set up: tree level}

Let us start by reviewing the predictions for the Cosmological Attractors
at tree level
\cite{Galante:2014ifa,Ferrara:2013rsa,Kallosh:2013yoa,Kallosh:2013tua,Giudice:2014toa}.
The Lagrangian of the models considered can be written as (with the
Planck mass set to one)
\begin{equation}
\mathcal{L}=\sqrt{-g}\left[\frac{1}{2}R-\frac{1}{2}K(\rho)(\partial\rho)^{2}-V(\rho)\right],
\label{eq:tree level}
\end{equation}
with
\begin{equation}
K=\frac{a_{p}}{\rho^{p}}+\frac{a_{p-1}}{\rho^{p-1}}+...,\,\,
V=V_{0}(1+c\rho+c_{2}\rho^{2}+..)
\label{eq:kin and pot}
\end{equation}
where $\rho\ll1$ is the parameter identifying the inflationary
regime.\footnote{$\rho\rightarrow1$ towards the end of inflation. Thus $\rho$ is defined in order for $c$ to be negative in \eqref{eq:kin and pot}.} $V$ is the tree level potential in the Einstein frame (in the
previous section labeled with $V^{(0)}$), while $V_0 = V|_{\rho =0}$
is the coupling dependent part of it.\footnote{The reason for these choices is to adopt the same
notation as \cite{Galante:2014ifa}.}

To first approximation the following happens
\cite{Galante:2014ifa}: the slow roll parameter $\eta$, and
consequently the spectral index, is completely determined by the order
of the leading pole in the kinetic term $(p)$; for $p=2$ the
tensor-to-scalar ratio $r$ will depend only on the residue of this
leading pole.\footnote{This approach is robust under
  perturbation of the non-canonical kinetic term $K$ with terms of one order higher in the leading pole, i.e. $K\subset a_{p+1}/\rho^{p+1}$
  \cite{Broy:2015qna}. }  We will now show this explicitly.

The first and second slow roll parameters are
\begin{equation}
\epsilon=\frac{1}{2}\left(\frac{V_{\rho}}{V}\right)^{2}K^{-1}
=\frac{1}{2}\left(\frac{V_{\rho}}{V}\right)^{2}\frac{\rho^{p}}{a_{p}},
\end{equation}
and
\begin{equation}\begin{split}\label{eta}
\eta&=\frac{V_{\chi\chi}}{V}=
-\frac{V_{\rho}}{V}\frac{d^{2}\chi}{d\rho^{2}}\left(\frac{d\chi}{d\rho}\right)^{-3}+\frac{V_{\rho\rho}}{V}\left(\frac{d\chi}{d\rho}\right)^{-2}\\[1mm]
&=\frac{p}{2}\frac{\rho^{p-1}}{a_{p}}c+O(\rho^{p}).
\end{split}\end{equation}
Here $\chi$ labels the canonical field defined via
$d\chi/d\rho=K^{\frac{1}{2}}$.  Further we introduced the notation
$V_\rho=d V/d\rho$, and likewise for higher derivatives.
From the previous expression we see that in all these models $\eta\gg\epsilon$ (one
order in $\rho$ difference). The number of efolds is given by
\begin{equation}
N\simeq\int^{\rho_{\star}}\left(\frac{V}{V_{\rho}}\right)Kd\rho
\simeq\frac{a_{p}}{c\rho_{\star}^{p-1}(1-p)}, \label{eq:N at tree}
\end{equation}
which implies 
\begin{equation}
\rho_{\star}=\left[\frac{Nc(1-p)}{a_{p}}\right]^{\frac{1}{1-p}}.
\end{equation}
Evaluating the slow roll parameters at horizon exit then gives
\begin{equation}\begin{split}
\eta_{\star}&\simeq\frac{p}{2(1-p)}\frac{1}{N}\,\,,\\[1mm]
\epsilon_{\star}&\simeq\frac{1}{2}c^{\frac{p-2}{p-1}}a_{p}^{\frac{1}{p-1}}\left(\frac{1}{(1-p)N}\right)^{\frac{p}{p-1}}.\label{eq:slwo roll}
\end{split}\end{equation}
For $p=2$ all dependence on the potential drops from the inflationary
parameters,
which become
\begin{equation}
n_{s}\simeq1-\frac{2}{N}\,\,,\quad\ r\simeq\frac{8a_{2}}{N^{2}}.\label{eq:predictions}
\end{equation}

\subsection{General set up: Quantum corrections\label{sub:General-set-up:}}

As we discussed in section \ref{sub:RG-improving-and}, even if the
inflaton field has a non-canonical kinetic term, the net effect of
considering leading order quantum corrections is captured by
substituting in the tree level action each coupling by its running
counterpart, i.e. $\lambda_{i}\rightarrow\lambda_{i}(t)$, modulo a
proper choice of the RG time $t=\tilde{t}$ (the one solving
\eqref{eq:sbar}). Thus, we consider the RG improved version of
\eqref{eq:tree level}, given by\footnote{In computing the effective action one usually expands around a constant background $\rho=\bar{\rho}$ where the tree level action is analytic. Even if we are in the limit of small $\rho$ we consider an expansion around a small but finite $\bar{\rho}$ where the action is analytic.}
\begin{equation}
\frac{\mathcal{L}_{{\rm eff}}}{\sqrt{-g}}\simeq-
\tfrac{1}{2}K(\rho,\lambda_{i}(\tilde{t}(\rho)))(\partial\rho)^{2}
-V(\rho,\lambda_{i}(\tilde{t}(\rho)).\label{eq:running action}
\end{equation}
It is worth making a remark here. The previous action captures the leading quantum correction as long as the effective action satisfies a Callan-Symanzik equation as \eqref{callan}, i.e. the theory is perturbatively renormalizable during inflation. For the case of interest ($p=2$) this can be seen in two ways. In terms of the canonical field, $K^{\frac{1}{2}}\partial\rho=\partial\chi$, $\rho\propto e^{-\chi/M}$, the action has an approximate shift symmetry, i.e. for $\chi\rightarrow \infty$ the action is invariant under $\chi\mapsto\chi+\rm{cost}$. This implies that all the divergences are proportional to $ e^{-\chi/M}$ and the counterterms will organize into a series of the same form as the tree level potential $V=V_{0}\sum_{n=1}^{\infty}c_{n}\rho^{n}=V_{0}\sum_{n=1}^{\infty}c_{n}e^{-n\chi/M}$. Thus we have at every order a finite number of counterterms. In terms of the non-canonical field $\rho$ the approximate shift symmetry turns into a scale symmetry of the action, $\rho\mapsto(\rm{cost})\rho$, in the limit of $\rho$ going to zero.\footnote{In terms of the non-canonical field $\rho$ this symmetry prevents that loop corrections generate higher inverse powers of $\rho$ in the non-canonical kinetic term.}

Let us now compute the effect of the quantum corrections on the inflationary parameters.   
For a potential of the form \eqref{eq:kin and pot}, the dependence on
the couplings is in $V_{0}$. The derivative of the potential with
respect to $\rho$, denoted by $V_{\rho}$, then becomes\footnote{Note
  that for the set-ups considered, even if the coefficients $c_{i}$
  had a dependence on the couplings that would only give contributions
  to the derivative that are  higher order in $\rho$, of the form
  $\sim
  V_{0}\left(\rho\frac{dc}{dt}\frac{d\tilde{t}}{d\rho}+..\right)$. }
\begin{equation}
V_{\rho}=V_{0}\sum_{n=1}^{\infty}nc_{n}\rho^{n-1}+\frac{\beta_{V_{0}}}{V_{0}}\frac{d\tilde{t}}{d\rho}V\label{eq:Vrho}
\end{equation}
where $c_{1}\equiv c$ (to match the tree level notation \eqref{eq:kin
    and pot}). At leading order we have
\begin{equation}
\frac{V_{\rho}}{V}\simeq c+\frac{\beta_{V_{0}}}{V_{0}}\frac{d\tilde{t}}{d\rho}.
\end{equation}
Now a key point of the argument kicks in. On the weak assumption
that $\tilde{t}$ can be simply expanded in a Taylor series about
zero, 
\begin{equation}
\frac{d\tilde{t}}{d\rho}=\sum_{k=0}^{\infty}d_{k}\rho^{k}\,\,,\label{eq:dt/drho}
\end{equation}
where the coefficients $d_{k}$ can depend implicitly on $\rho$,
we have 
\begin{equation}
\frac{V_{\rho}}{V}=c+\frac{\beta_{V_{0}}}{V_{0}}d_{0}+O(\rho).
\label{eq:VrhoV}
\end{equation}
Thus, the effect of the RG flow will be only a rescaling of the factor
\begin{equation}
c\rightarrow\mathcal{C}\equiv c\left(1+\frac{\beta_{V_{0}}}{V_{0}}\frac{d_{0}}{c}\right).\label{eq:c'}
\end{equation}
As we show now, this will be the only relevant effect which has no
consequences for the inflationary parameters.

Let us compute the number of efolds. The leading term in the integrand
is the same as \eqref{eq:N at tree} with the replacement \eqref{eq:c'}.
On the other hand, the factor $a_{2}/\mathcal{C}\equiv D$ is not
a constant anymore and it depends implicitly on $\rho$. Expanding
it in a Taylor series about $\rho_{\star}$ gives
\begin{align}
N&\simeq\int^{\rho_{\star}}\frac{d\rho}{\rho^{2}}\frac{a_{2}}{\mathcal{C}}
\nonumber\\
&=\int_{\rho_{\star}}\frac{d\rho}{\rho^{2}}\left(D_{\star}+\beta_{D\star}\frac{d\tilde{t}}{d\rho}(\rho-\rho_{\star})+..\right).\label{eq:NN}
\end{align}
Given the assumption \eqref{eq:dt/drho}, we observe that $D$ can be
considered constant over the integration domain within our
approximation, i.e. $D\simeq D_{\star}$. In fact, all the other terms,
starting from the second one within the brackets, give contributions
that are at most of order $\sim\ln\rho_{\star}$, which gives an order higher in the $1/N$ expansion. These contributions enter at the same
order as the corrections from the subleading poles in the kinetic
term, which were already neglected at tree level. The number of efolds
then becomes\footnote{We will comment on the possibility of quantum corrections flipping the sing of $\mathcal{C}_{\star}$ in the next section.}
\begin{equation}
N\simeq-\frac{a_{p\star}}{\mathcal{C}_{\star}\rho_{\star}},
\label{eq:N running}
\end{equation}
which is simply \eqref{eq:N at tree} with $p=2$ and the couplings
(which now are not constant anymore) evaluated at the horizon crossing
$\rho_{\star}$. Therefore $\epsilon_{\star}$ will be exactly the same
as the tree level expression, but with $c$ replaced by
$\mathcal{C}_{\star}$ (whose dependence drops out for $p=2$). 
From \eqref{eq:N running} we note that the quantum corrections, encoded in   $\mathcal{C}_{\star}$, generically will not prevent this model from generating a large enough number of efolds. In fact $\mathcal{C}_{\star}$ can be in principle arbitrarily large, this will simply imply a shift of $\rho_{\star}$ towards smaller values in order to get the desired number of efolds. 
In computing $\eta$, one has to be a little more careful since an extra
contribution could come from the derivative of the non-canonical
kinetic term in \eqref{eq:running action}.  In fact, in (\ref{eta}),
we should consider
\begin{equation}
\frac{dK^{\frac{1}{2}}}{d\rho}\equiv\frac{d^{2}\chi}{d\rho^{2}}=\frac{1}{2}\rho^{\frac{p}{2}}a_{p}^{\frac{1}{2}}
\left(\!\!-p\rho^{-p-1}+\frac{\beta_{a_{p}}}{a_{p}}\frac{d\tilde{t}}{d\rho}\rho^{-p}\!\right)\Big|_{p=2}\!.
\end{equation}
However, as long as \eqref{eq:dt/drho} is satisfied, the second term
in this expression gives higher order contributions to $\eta$.
Thus  (\ref{eta}) becomes 
\begin{equation}
\eta=-\frac{V_{\rho}}{V}\frac{d^{2}\chi}{d\rho^{2}}\left(\frac{d\chi}{d\rho}\right)^{-3}+O(\rho^{2})=\rho\frac{\mathcal{C}}{a_{2}}+O(\rho^{2}),
\label{eq:etabeta}
\end{equation}
which is the tree level expression at leading order with again $\mathcal{C}$
playing the role of $c$. It is then obvious that inverting \eqref{eq:N running}
and substituting $\rho_{\star}$ in the the slow roll parameters gives
the same \eqref{eq:predictions} for $n_{s}$ and $r$,
\begin{equation}
n_{s}\simeq1-\frac{2}{N}\,\,,\quad r\simeq\frac{8a_{2\star}}{N^{2}}.\label{eq:imppred}
\end{equation}
Summarizing, the effect of the RG flow enters in three ways.  First,
in the efolds dependence of $\rho_{\star}=\rho_{\star}(N)$.  Second,
by giving a rescaling $c\rightarrow\mathcal{C}$ in the slow roll
parameters (before evaluating them at the horizon crossing) and third
from the extra contribution to the derivative of $K$ in
$\eta$. Nevertheless, if the condition \eqref{eq:dt/drho} is
satisfied, this latter gives simply higher order contributions, while
the first and second points compensate each other. Different running
histories (encoded in $(\beta_{V_{0}}/V_{0})_{\star}$ in
$\mathcal{C}_{\star}$) will just imply a different value of the field
at the horizon exit $\rho_{\star}$.  This effect cancels with the
shifted expressions of the slow roll parameters.\footnote{Suppose that
  \eqref{eq:dt/drho} is not satisfied, for example
  $\tilde{t}\propto\frac{k}{\rho}$.  In the number of efolds the
  second term in \eqref{eq:NN} will now give
  $\int\frac{d\rho}{\rho^{2}}\beta_{D\star}\frac{d\tilde{t}}{d\rho}(\rho-\rho_{\star})=\frac{k}{\rho_{\star}}\beta_{D\star}+{\rm
    h.o.}$,
  which is of the same order as the leading term
  $D_{\star}/\rho_{\star}$, and thus \eqref{eq:N running} and the
  relation $\rho_{\star}(N)$ is altered. As a consequence, the leading
  order slow roll parameters at horizon exit will depend on the beta
  functions.}

Note that the arguments presented are valid as long as the quantum
corrections encoded in $\mathcal{C}$ do not break the perturbative
expansion in $\rho$. In
general $|\mathcal{C}|\sim O(1)$ and \eqref{eq:imppred}
follows. Consider the term between brackets in \eqref{eq:c'}, denoted
by $F$, i.e.
\begin{equation}
\mathcal{C}=c\left(1+\frac{d_{0}}{c}\frac{\beta_{V_{0}}}{V_{0}}\right)\equiv cF,\label{eq:F}
\end{equation}
if $F$ is positive during inflation, then
${\rm sign}[\mathcal{C}_{\star}]<0$ and the predictions will be the
same as for the tree level case. 

\begin{figure}
\includegraphics[width=0.35\paperwidth]{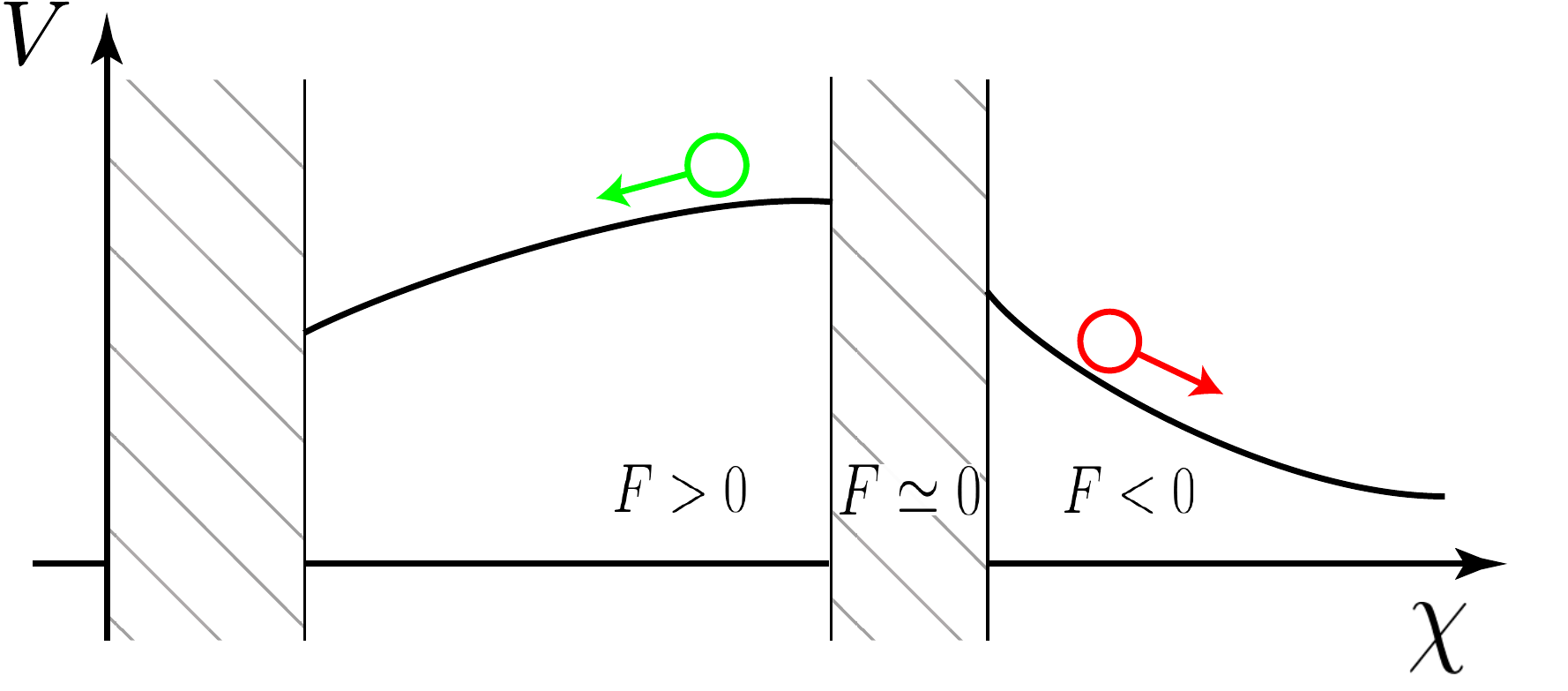} \caption{Possible features of the effective potential due to RG effects.\label{fig:Potential-feature-due}}

\end{figure}

\subsection{
Maximum and breakdown of perturbativity
}\label{sub:common}

  The conclusion that the tree level predictions are not affected by
  the RG corrections is valid as long as the perturbative expansion in $\rho$ holds and inflation takes place on the plateau of the potential. It may be that the potential
  develops an extremum because of the running; this is purely a
  quantum effect in that the tree level potential has no extremum in
  the inflationary regime. For fine-tuned parameters it is then
  possible to obtain inflation near the maximum or inflection
  point. In this case the details of the inflationary scenario
  will depend sensitively on the quantum corrections.

To see the appearance of an extremum in the potential, consider
  its slope. For $|F| \gtrsim \rho$ the perturbative expansion is
  valid and $V_\rho = V_0 c F +\O(\rho)$, see \eqref{eq:VrhoV}.  It
  follows that going from a region in field space with
  $F \gtrsim \rho$ to a region with $F \lesssim -\rho$, the slope of
  the potential has changed sign. This can only happen if there is a
  (at least one) maximum in between. In general, one cannot calculate
  the location of the maximum analytically though, as the perturbative
  expansion breaks down exactly in the in between region where
  $F \sim 0$.

  For the particular choice of normalization scale
  $\bar{\mu}(\tilde{t})\propto V^{\frac{1}{4}}$ the slope of the
  potential factorizes in a classical piece times a quantum correction
  at all orders, see \eqref{eq:Vrho2} below.  Then it can actually be shown
  analytically that the regime where perturbativity breaks down
  coincides with the development of a maximum in the effective
  potential.  This choice of normalization scale is appropriate for
  Higgs inflation and it also appears generically in the $\alpha$ and
  $\xi$-attractor models considered. It can be parametrized
  $\bar{\mu}(\tilde{t})=(V_0 \mu_{0})^{1/4}(1+c\rho+..)^{1/4}$, where
  $V_0$ and $\mu_{0}$ depend explicitly on the couplings but do not
  explicitly depend on the field $\rho$. It is not hard to see that
this satisfies \eqref{eq:dt/drho}.  The RG time is then
\begin{equation}
\tilde{t}=\ln\bar{\mu}(\tilde{t})=\frac{1}{4}\ln\left(\mu_0V\right).
\label{eq:t=00003Dpotential}
\end{equation}
Taking the derivative with respect to $\rho$ gives 
\begin{equation}
\frac{d\tilde{t}}{d\rho}=\frac{1}{4}\left(\frac{V_{\rho}}{V}+\frac{\beta_{\mu_0}}{\mu_0}\frac{d\tilde{t}}{d\rho}\right)\implies\frac{d\tilde{t}}{d\rho}=\frac{V_{\rho}/4V}{1-\beta_{\mu_0}/4\mu_0},
\label{eq:dtdrho2}
\end{equation}
which allows us to write an exact expressions for $V_\rho$ without
needing to solve \eqref{eq:Vrho} iteratively. Inserting the previous
expression in \eqref{eq:Vrho} gives 
\begin{equation}
V_{\rho}=V_{0}\left(\sum_{n=1}^{\infty}nc_{n}\rho^{n-1}\right)\frac{\left(1-\frac{\beta_{\mu_0}}{4\mu_0}\right)}{\left(1-\frac{\beta_{\mu_0}}{4\mu_0}-\frac{\beta_{V_{0}}}{4V_{0}}\right)},\label{eq:Vrho2}
\end{equation}
 and thus\footnote{Note that \eqref{eq:F} and \eqref{eq:bb} are in agreement. Combining
\eqref{eq:dtdrho2} and \eqref{eq:Vrho2}, $d\tilde{t}/d\rho$ can
be written in the form \eqref{eq:dt/drho} with
$d_{0}=({c}/{4})(
  1-\frac{\beta_{\mu_0}}{4\mu_0}-\frac{\beta_{V_{0}}}{4V_{0}}
 )^{-1}$.
Now using \eqref{eq:F} gives for $F$ the same expression
as in \eqref{eq:bb}.} 
\begin{equation}
\mathcal{C}=c\frac{\left(1-\frac{\beta_{\mu_0}}{4\mu_0}\right)}{\left(1-\frac{\beta_{\mu_0}}{4\mu_0}-\frac{\beta_{V_{0}}}{4V_{0}}\right)}\equiv cF.\label{eq:bb}
\end{equation}
Now switching to the canonical field $\chi$, we thus find that for
$F=0$ the slope vanishes
$V_{\chi}=-K^{-\frac{1}{2}}F(c\rho+O(\rho^2))=0$.\footnote{Here we
  used that $d\rho/d\chi=-K^{-\frac{1}{2}}<0$ since $K>0$ and
  $\rho\rightarrow0$ for $\chi\rightarrow\infty$.} The inflaton
potential develops an extremum and $\epsilon$ is identically zero at
all orders. Approaching this point, when $F\simeq\rho$, our
perturbative analysis breaks down. To show that the potential develops
a  maximum consider the curvature at the extremum 
$V_{\chi\chi}|_{F=0}
=K^{-1}F_{\rho}c(1+O(\rho))$.
As $c$ is negative, it follows that 
$\rm{sign}[V_{\chi\chi}]|_{\rho_{\rm
    ext}}=-\rm{sign}[F_{\rho}]|_{\rho_{\rm ext}}$ where $\rho_{\rm
  ext}$ is the field value at the extremum.
 Now since
$F(\rho)>0$ for $\rho>\rho_{\rm ext}$, we have
$F_{\rho}(\rho_{\rm ext}) \geq 0$. This implies that
$\rm{sign}[V_{\chi\chi}]|_{\rho_{\rm ext}}<0$, i.e. the extremum is indeed
a maximum (or an inflection point for double fine-tuned parameters
\cite{critical1, critical2, Enckell:2016xse}). Note that, also this
result is independent on the particular $\beta$-functions.


One can contemplate the possibility of inflation happening on the
other side of the maximum.  $F$ is negative here as
${\rm sign}[\mathcal{C}_\star]$ is reversed. However, this describes a completely different kind of
inflation with the inflaton rolling on the other side of the maximum
(see Fig.  \eqref{fig:Potential-feature-due}). Therefore, when a
maximum develops due to quantum corrections, one has to assume initial
conditions such that inflation starts with the inflaton always on the
``correct side'' of the maximum. In Higgs inflation this is equivalent
to the observational request to end up in the electroweak vacuum after
inflation.  For the wider class of models considered here, the
assumption is still reasonable since the vacuum at the origin is tuned
to have zero (small) cosmological constant and this is where inflation
is assumed to end; the minimum at large field values might not only be
large (negative or positive), but also in the regime where any
calculational control is lost as quantum gravity corrections may be
large.

\subsection{Higher orders}
\label{sec:higher}

Even if the arguments shown in section \ref{sub:RG-improving-and} are
general, it is still not clear if the result presented in
section \ref{sub:General-set-up:} hold beyond leading order (LO). Consider the
RG improved $L$-loops potential, i.e.
$V=V_{0,{\rm eff}}(\tilde{t})(1+..)$ where
$V_{0,{\rm eff}}=V_{0}(\tilde{t})+V_{0}^{1}(\tilde{t})+..$ only depend explicitly
on the couplings. Since the cancellation of the RG effects in the
inflationary predictions does not depend explicitly on the particular
$V_{0}$ nor on the order of $\beta_{V_{0}}$, everything still follows
replacing $V_{0}\rightarrow V_{0,{\rm eff}}$ and
$\beta_{V_{0}}\rightarrow\beta_{V_{0,{\rm eff}}}$. However, including
higher loops contributions we can in general no longer absorb the
effect of the anomalous dimensions in the canonical field, as was
discussed at the end of section \ref{sub:RG-improving-and}. This may not necessarily hold beyond LO, for which further investigations would be required.

\section{Applications}

We now discuss the classes of models that can be written in the
general form \eqref{eq:kin and pot}; these are the $\alpha$- and
$\xi$- attractors. As a particular case of this latter we first
consider Higgs inflation (HI) \cite{bezrukov1}.
Here the condition \eqref{eq:dt/drho} on the renormalization scale is
determined by couplings of the Higgs to the other Standard Model
particles. Nothing guarantees beforehand that this is still valid for the more general class of models considered. Nevertheless, in \ref{sub:Renormalization-scale-for} we discuss how the weak condition (\ref{eq:dt/drho}) holds naturally also
for the Cosmological Attractors. 

It thus follows that the tree level predictions for all these models are robust against quantum corrections.


\subsection{Higgs Inflation\label{sub:Special-case:-Higgs}}

The argument outlined in the previous section applies to Higgs
inflation. This is the reason why in
\cite{Fumagalli:2016lls} it was found that the dependence on the
$\beta$-functions drops out of the inflationary predictions. In fact,
the kinetic term and the potential for Higgs inflation (in the Einstein
frame and in unitary gauge) can be written as a Laurent series
in $\rho=\Omega^{-1}=1/(1+\xi\phi^{2})$ as
\begin{align}
K&=\frac{3}{2\rho^{2}}+\frac{1}{4\xi(1-\rho)\rho^{2}}
\nonumber\\
&\simeq\frac{3}{2}\left(1+\frac{1}{6\xi}\right)\frac{1}{\rho^{2}}+\frac{1}{4\xi}\sum_{i=-1}^{\infty}\rho^{i}
\end{align}
and 
\begin{equation}
V=V_{0}(1-2\rho+\rho^{2})\label{eq:treeHI}
\end{equation}
with $V_{0}={\lambda}/{(4\xi^{2})}$. This is exactly of the general form 
\eqref{eq:kin and pot} with $p=2,\,c=-2$ and $a_{2}=3/2(1+1/6\xi)$.
The slow roll parameters 
at tree level are given by \eqref{eq:slwo roll},
which implies 
\begin{equation}
n_{s}\simeq1-\frac{2}{N}, \quad r\simeq12\left(1+\frac{1}{6\xi}\right)\frac{1}{N^{2}}.
\end{equation}
Large values of $\xi$ are needed to fit the power spectrum of the
scalar perturbations. In this limit the $\xi$-dependence disappears 
form the tensor-to scalar ratio.

Let us now turn to the quantum corrections. In Higgs inflation there
is a natural choice for the RG time $\tilde{t}$, which is chosen such
that it minimizes the largest logs in the Coleman-Weinberg potential
in agreement with \eqref{eq:sbar}.  If Higgs
inflation is embedded in the Standard Model, the dominant quantum
corrections come from the $W$ and $Z$ bosons and from the top quark
masses \cite{bezrukov_loop}. Their masses all scale the same way with the Higgs field,
namely as 
$M_{W}=gf(\phi)/2,\,\,\,M_{Z}=(g^{2}+g'^{2})^{\frac{1}{2}}f(\phi)/2,\,\,\,M_{t}=y_{t}f(\phi)/\sqrt{2})$
with $g_{1},g_{2},y_{t}$ the $U(1)$, $SU(2)$ and Yukawa couplings
respectively and $f(\phi)=\phi/\Omega^{\frac{1}{2}}$. Using the
notation of section \ref{sub:RG-improving-and}, we choose for
simplicity
\begin{equation}
s=\ln\left(\frac{f(\phi)}{\mu}\right).
\end{equation}
The logs is minimized for $\bar{s}(\tilde{t})=0$ which implies
\begin{equation}\begin{split}
\tilde{t}=\ln\frac{\phi}{(1+\xi(\tilde{t})\phi^{2})^{\frac{1}{2}}}=\frac{1}{4}\ln\left(\frac{4V(\tilde{t})}{\lambda(\tilde{t})}\right).
\label{eq:s_HI}
\end{split}\end{equation}
To get the last expression it was used that the classical potential can be written
as $V=\lambda f^{4}(\phi)/4$.  We have given the masses and
renormalization scale in the Einstein frame, as this is where the
inflationary observables are most easily computed.  We note however
that physics is frame independent; even if initially the loop
corrections and renormalization scale is computed in the Jordan frame,
and only afterwards the results are transformed to the Einstein frame,
this would give the same result for the renormalization scale
\eqref{eq:s_HI} \cite{damien,Fumagalli:2016lls}.

It follows that $\tilde{t}$ is actually of the form
\eqref{eq:t=00003Dpotential} with $\mu_0=4/\lambda$ which
satisfies the generic assumption \eqref{eq:dt/drho}. Using
\eqref{eq:dtdrho2}, \eqref{eq:Vrho2} and \eqref{eq:bb} with
$\beta_{\mu_0}/\mu_0=-\beta_{\lambda}/\lambda$ and
$\beta_{V_{0}}/V_{0}=\beta_{\lambda}/\lambda-2\beta_{\xi}/\xi$
gives\footnote{This can be matched to the notation used in
  \cite{Fumagalli:2016lls}, where the small parameter
  $\delta=(\xi\phi^{2})^{-1}$ was used as expansion parameter.  In
  that notation
  $ \rho=\delta(\delta+1)^{-1}\simeq\delta+O(\delta^{2})$ and
  $
  \phi=\left(\frac{1-\rho}{\xi\rho}\right)^{\frac{1}{2}}$,
  which gives
\begin{equation}\nonumber
\frac{dt}{d\phi}=\frac{\xi^{\frac{1}{2}}\delta^{\frac{3}{2}}}{1+\frac{\beta_{\xi}}{2\xi}+\delta},\,
  \frac{d\phi}{d\rho}=-\frac{1}{2\xi^{\frac{1}{2}}}\frac{\rho^{-\frac{3}{2}}}{(1-\rho)^{\frac{1}{2}}}
+\frac{\beta_{\xi} (1-\rho)^{\frac{1}{2}}}{\xi^{\frac{3}{2}}\rho^{\frac{1}{2}}}\frac{dt}{d\rho},
\end{equation}
It then follows that at leading order
$\frac{dt}{d\rho}=\frac{dt}{d\phi}\frac{d\phi}{d\rho}$
agrees with \eqref{eq:dt_HI}.}
\begin{equation}
\frac{d\tilde{t}}{d\rho}=-\frac{1}{2}\frac{1}{\left(1+\frac{\beta_{\xi}}{2\xi}\right)}+O(\rho)
\label{eq:dt_HI}
\end{equation}
and
\begin{equation}
\mathcal{C}=c\left(1+\frac{\beta_{V_{0}}/4V_{0}}{1-\frac{\beta_{\mu_0}}{4\mu_0}}\right)=-2\left(\frac{1+\frac{\beta_{\lambda}}{4\lambda}}{1+\frac{\beta_{\xi}}{2\xi}}\right)\equiv-2F.
\end{equation} 
Since Higgs inflation is a specific example of the general set-up
considered in section \ref{sub:General-set-up:} the shift of $C$
by the quantum corrections drops out of the inflationary predictions,
which are equal to the tree level results.  This conclusion holds as
long as the perturbative expansion in $\rho$
is valid. As discussed in \ref{sub:common} for
$F_{\star}\simeq \rho_{\star}$
the potential develops a maximum, which for fine-tuned parameters can
be used for ``hilltop inflation'' in agreement with the CMB data. This
possibility was studied numerically in \cite{Fumagalli:2016lls, Enckell:2016xse}.

Our choice of renormalization scale \eqref{eq:s_HI} is referred to as
``prescription I'', see \cite{damien} and section 2.3 of
\cite{Fumagalli:2016lls} for an extensive discussion. This is the
natural choice since it minimizes the log preserving the asymptotic
shift symmetry of the potential \cite{shap,Fumagalli:2016lls}. In the
literature "prescription II" has been also considered as well, defined by
\begin{equation}
\tilde{t}=\ln\phi=\frac{1}{2}\ln\left[\frac{1}{\xi}\left(\frac{1-\rho}{\rho}\right)\right].
\label{eq:prescription2}
\end{equation}
This corresponds to a different UV completion of the theory (where the
potential is already altered at tree level in the large field
regime). Using this prescription for the renormalization scale we
can immediately see that the previous cancellation does not take place
anymore. In fact
$\frac{dt}{d\rho}=-\frac{1}{2}\frac{1}{\rho(1-\rho)}$
which is not of the form \eqref{eq:dt/drho}. This is the reason why in
\cite{bezrukov_loop,kyle,cliffnew} (see also \cite{Marzola:2016xgb}), where prescription II was
considered, features for $n_{s}$ and $r$ have been observed.\footnote{For examples where the potential is altered explicitly at tree level see \cite{Salvio:2015kka,Barbon:2015fla}.}

\subsection{$\alpha$-attractors}

The $\alpha$-attractors \cite{Ferrara:2013rsa,Kallosh:2013yoa},which
are a generalization of conformal attractors
\cite{Kallosh:2013hoa,Kallosh:2013daa}, are described by the
Lagrangian
\begin{equation}
\frac{\mathcal{L}}{\sqrt{-g}}=R-\frac{1}{2}K(\partial\phi)^{2}-V(\phi)\label{eq:alpha action}
\end{equation}
 with 
\begin{equation}
K=\frac{\alpha}{(1-\phi^{2}/6)^{2}}\,\,,\quad V=\alpha f^{2}(\phi/\sqrt{6}).
\end{equation}
The Starobinsky model \cite{Starobinsky:1980te} also belongs to this class for a particular choice of $f$ and $\alpha=1$.  
Through the change of variable 
\begin{equation}\label{phirho}
\frac{\phi}{\sqrt{6}}=\frac{1-\rho}{1+\rho},
\end{equation}
the inflaton Lagrangian becomes
\begin{equation}
\frac{\mathcal{L}}{\sqrt{-g}}=R-\frac{1}{2}\left(\frac{3\alpha}{2\rho^{2}}\right)(\partial\rho)^{2}-\alpha
f^{2}\left(\frac{1-\rho}{1+\rho}\right).
\label{eq:apha attractors}
\end{equation}
From \eqref{eq:slwo roll} it then follows that for quite
generic\footnote{The function $f$ should not be singular at $\phi=1$, or
  equivalently at $\rho=0$, such that
it is possible to expand it as $f=V_{0}(1+c\rho+O(\rho^{2}))$.}
 $f$ the tree level results are 
\begin{equation}
n_{s}\simeq1-\frac{2}{N}\,\,,\quad r\simeq\frac{12\alpha}{N^{2}}.\label{eq:alpha predictions}
\end{equation}
As we will motivate in section \ref{sub:Renormalization-scale-for}
the RG time can be chosen as in \eqref{eq:t=00003Dpotential}. The
analysis including quantum corrections is then a special case of the
general discussion in section \ref{sub:General-set-up:}. As was shown
there, the inflationary observables are not affected by quantum
corrections 
as long as
$F$ does not break the perturbative expansion in powers of $\rho$.

\subsection{$\xi$-attractors}

The $\xi$-attractors are models in which the inflaton is non-minimally coupled to the gravity sector \cite{Kallosh:2013tua,Giudice:2014toa,Csaki:2014bua}. They are described by a Lagrangian of the form 
\begin{equation}
\frac{\mathcal{L}}{\sqrt{-g}}=\frac{\Omega}{2}R-\frac{1}{2}K_{J}(\partial\phi)^{2}-V_{J}(\phi).\label{eq:xi attractors}
\end{equation}
After the usual conformal transformation of the metric $g=\Omega^{-1}g_{E}$,
the gravity sector is in the standard form and the Einstein frame
field metric and potential are
\begin{equation}
K=\frac{K_{J}}{\Omega}+\frac{3}{2}\frac{\Omega'^{2}}{\Omega^{2}}\,\,,\quad V=\frac{V_{J}}{\Omega^{2}}.\label{eq:special attractors}
\end{equation}
Let us briefly review which classes of models belong to the $\xi$-attractor
family. In \cite{Galante:2014ifa} it has been shown that for the
special choice $K_{J}=\frac{1}{4\xi}\frac{\Omega'^{2}}{\Omega},\quad V_{J}=\Omega^{2}U(\Omega)$
(\textsl{special attractors}) the models are completely equivalent
to the $\alpha$-attractors with the identification $1+\frac{1}{6\xi}\equiv\alpha$.
In fact, with this choice of $K_{J}$ in \eqref{eq:special attractors},
the Einstein frame field space metric
$K$ becomes exactly the one of \eqref{eq:apha attractors}. Other
subclasses of $\xi$-attractors are the \textsl{induced inflation models
} \cite{Giudice:2014toa} described by
\begin{equation}
\Omega=\xi f(\phi)\,,\quad K_{J}=1\,,\quad V_{J}=V_{0}(\Omega-1)^{2},
\end{equation}
and the condition that $\Omega\rightarrow0$ as $\phi\rightarrow0$; the
\textit{universal attractors} \cite{Kallosh:2013tua} satisfy
$\Omega\rightarrow1$ as $\phi\rightarrow0$, with
\begin{equation}\label{omega}
\Omega=1+\xi f(\phi)
\end{equation}
and the same $V_{J}$ and $K_{J}$ as the induced inflation
models. Higgs inflation is a particular example of a universal
attractor model. For $\xi\rightarrow\infty$ all these models give
classically the same predictions at leading order in $N^{-1}$, which
coincide with the predictions of the $\alpha$-attractors for
$\alpha\rightarrow1$. Indeed, in this limit the first term in $K$
in \eqref{eq:special attractors} can be neglected. Thus the Lagrangian
in the Einstein frame, after the field redefinition
$\rho=\Omega^{-1}$, becomes
\begin{equation}
\frac{\mathcal{L}}{\sqrt{-g}}\simeq\frac{R}{2}-\frac{1}{2}\left(\frac{3}{2\rho^{2}}\right)\left(\partial\rho\right)^{2}-V_{0}(1-\rho)^{2}.\label{eq:xi attr in rho}
\end{equation}
This is of the form \eqref{eq:tree level}-\eqref{eq:kin and pot} with
the leading pole of order two in the kinetic term ($p=2$). Therefore
the predictions coincide at first order with \eqref{eq:alpha
  predictions} in the $\alpha\rightarrow1$ limit and given
\eqref{eq:dt/drho} the conclusions on the RG flow corrections are the
same as in the previous sections.   In the next section we further
discuss the choice of renormalization scale.

\subsection{Renormalization scale for Cosmological Attractors\label{sub:Renormalization-scale-for}}

Whereas the choice of $\tilde{t}$ in Higgs inflation is determined by
the known couplings of the Higgs to the Standard Model fields, it is
not clear a priori whether $\tilde{t}$ for a generic $\alpha$ or $\xi$
attractor satisfies \eqref{eq:dt/drho}. This depends on the inflaton
couplings to the other fields, which is model dependent. Let us
consider first the case where only the inflaton loops are relevant in
the inflationary regime (which happens when the coupling to all other
fields is suppressed).\footnote{The general argument in this context
  \cite{Kallosh:2016gqp} states that the one loop correction is
  negligible since it is suppressed by an extra factor $m_{\chi}^2$ as
  compared to the tree level potential.}
The one loop correction in the Einstein frame is then
of the form (remember that the canonical inflaton field is labeled
with $\chi$) \cite{Bilandzic:2007nb,George:2012xs,Markkanen:2013nwa}
\begin{equation}\label{1loop}
V^{(1)}\propto m_{\chi}^{4}\ln\left(\frac{m_{\chi}^2}{\mu^2}\right)\simeq m^4_{\chi}\ln\left(\frac{H^2}{\mu^2}\right)
\end{equation}
with the inflaton mass given by\footnote{In all these kind of models we can neglect the backreaction of gravity
since $\epsilon\ll\eta$ \cite{damien}.} 
\begin{equation}\begin{split}
m_{\chi}^{2}&\simeq V_{\chi\chi}-2H^{2}\simeq V\left(\eta-\frac{2}{3}\right)\\[1mm]
&\simeq-\frac{2}{3}V_{0}\left(1+c\left(1-\frac{3}{2a_{p}}\right)\rho+O(\rho^{2})\right).
\end{split}
\label{eq:mchi}
\end{equation}
Thus the RG time satisfying \eqref{eq:sbar} is
\begin{eqnarray}\label{relation}
\tilde{t} & \simeq & \ln\left(\bar{m}_{\chi}^{2}(\tilde{t})\right)\sim\ln(H^2)\sim\ln V,\label{eq:mxhi}
\end{eqnarray}
where with $\ln(..)\sim\ln V$ we mean that the arguments of the two logs contain
the same powers of $\rho$. This is enough to ensure that \eqref{eq:dt/drho}
is satisfied.
The reason for this is simple. If the theory is renormalizable
 in the EFT sense, the one loop term (as any other log term
in the effective potential)
 can be reabsorbed order by order
in the tree level part.\footnote{The slow roll parameter $\eta$ in
  terms of the canonical field is  $\eta\propto e^{-\chi/a_{2}^{\frac{1}{2}}.}$
The fact that we can reabsorb the $\eta$-dependent term in \eqref{eq:mchi}
is basically another way of saying that radiative corrections do not spoil the quasi shift symmetry
of the classical potential.}

Let us now consider the case where
the loop corrections are dominated by other fields running in the loops. We focus on the
inflaton $\phi$ coupled to a scalar field $\sigma$; the results
straightforwardly generalize to fermion and gauge fields.  Here $\phi$
is the original field appearing in the Lagrangians that the define the
models, \eqref{eq:alpha action} for the $\alpha$-attractors and
\eqref{eq:xi attractors} for the $\xi$-attractors.  In the small field
regime $K \simeq 1$ and $\phi\simeq\chi$, and the $\phi$ field is the
canonical renormalizable field.  If we demand the theory to be
renormalizable in the small field regime, the coupling between
$\sigma$ and $\phi$ has to be in a standard renormalizable form. This
automatically implies that a one loop term like \eqref{eq:mxhi}, with
$m_{\sigma}^{2}=\partial^{2}V(\phi,\sigma)/\partial\sigma^{2}$, can be
reabsorbed in the tree level potential in the inflationary
regime. Indeed, $m_{\sigma}^{4}$ and $V$ share the same field
dependence over the whole field regime. Thus, since the
renormalization scale accommodates a subset of the powers of $\phi$
contained in the classical potential it implies that, once rewritten
in terms of $\rho$, it will equally have a subset of the powers of
$\rho$ that are contained in $V$.

Let us illustrate the previous statements with a simple
example. Consider a coupling $\L \supset g^{2}\phi^{2}\sigma^{2}$ in
the action \eqref{eq:alpha action} \cite{Kallosh:2016gqp}. Assume for simplicity that the scalar field $\sigma$ has no bare mass term. In the small field regime the $1$-loop contribution will
be of the form $V^{(1)}\sim g^{4}\phi^{4}\ln(g^{2}\phi^{2}/\mu^{2})$,
which can be absorbed in a quartic tree level potential
$V\sim\lambda\phi^{4}$.  For $\alpha$-attractors, in the large field
regime $V^{(1)}$ and $V$ have exactly the same functional form, which
implies $\tilde{t}=1/4\ln(\bar{g}^{4}\phi^{4}(\rho))$. This written in
terms of $\rho$ using \eqref{phirho} will give a RG time $\tilde{t}$
satisfying \eqref{eq:dt/drho} (in particular, it will be of the form
\eqref{eq:t=00003Dpotential}).  A similar applies to the
$\xi$-attractors.  The coupling generates loop contributions that can
be reabsorbed in the Jordan frame potential $V_{J}$ over the whole
field range, fully analogously to the situation in the
$\alpha$-models. In the large field regime, once we transform to the
Einstein frame, all the mass scales (including the renormalization
scale, see \cite{Fumagalli:2016lls,damien} and \cite{Kannike:2015apa})
are rescaled as $m\rightarrow m/\Omega^{\frac{1}{2}}$. This still
ensures that $\tilde{t}$ is of the form \eqref{eq:dt/drho}; the
explicit example in this context is given by Higgs inflation in
section \ref{sub:Special-case:-Higgs}.

The argument presented so far comes with a caveat. We demanded the low
energy regime to be renormalizable, which is not necessary for a working
inflationary model.  On top, we can relax the constraint by only
asking the theory to be renormalizable in the EFT sense in the small
field regime; this opens the possibility that the theory is only
defined up to some cutoff scale $\Lambda$ in this regime. To be specific,
consider the $\alpha/\xi$-attractors actions  \eqref{eq:alpha action} and \eqref{eq:xi attractors} augmented with an interaction term of the form
\begin{equation}\label{log}
\mathcal{L}_{\rm{I}}= \Lambda^2g\sigma^2\ln\left(1+\frac{\phi^2}{\Lambda^2}\right).
\end{equation} 
Expanding in $\phi\ll\Lambda$ gives the first term of the previous
example plus a tower of higher order operators suppressed by powers of
the cutoff. The theory is clearly renormalizable in the EFT sense in
the small field regime. However, things might change in the large
field regime.

In the $\alpha$ models, using the expression of $\phi$ in terms of
$\rho$ given by \eqref{phirho}, we get
\begin{equation}
\mathcal{L}_{\rm{I}}\sim g\sigma^2\left(a_0+a_1\rho+O(\rho^2)\right),
\end{equation}
where $a_i$ are just the numerical coefficients of the
expansion. Therefore, even in the high field regime, the quantum
corrections generated by the field $\sigma$ (proportional to
$m_{\sigma}^4$) can be absorbed order by order in the tree level
potential. The RG time $\tilde t$ satisfies a relation like
\eqref{relation} and the main conclusions of section
\ref{sub:General-set-up:} still follow.  The same argument does not
apply to the $\xi$-attractors though. Here the interaction
term in the Einstein frame becomes
\begin{equation}
\mathcal{L}_{\rm{I}}\propto \frac{g\sigma^2}{\Omega^2}\ln\left(1+\frac{\phi^2}{\Lambda^2}\right)\simeq g\sigma^2 \rho^2 \ln\left(\rho^{-1}\right),
\end{equation}
where for simplicity we have taken $f(\phi)\propto \phi^2$ in
\eqref{omega}.\footnote{Similar results follow for generic $f(\phi)$ polynomial in $\phi$ and generic powers of $\phi$ appearing in the log in \eqref{log}.}This cannot be expanded in powers of $\rho$ as before, leading apparently\footnote{We say "apparently" because if the theory is not renormalizable even perturbatively, all the arguments of section \ref{sub:RG-improving-and} do not apply anymore and talking about RG improved action is misleading.} to a
choice for the renormalization scale which does not fulfill condition
\eqref{eq:dt/drho}.  The interaction 
renders the theory not
renormalizable in the EFT sense.  Hence, interaction terms like the
one considered here are simply not allowed on this ground --- it is
important to remember that our whole analysis in section
\ref{sub:RG-improving-and} is based on the assumption that inflation
is described by a perturbatively normalizable EFT.

\section{Conclusions}

In this letter we analyzed the Cosmological Attractor models using the
RG improved action to include the leading log quantum
corrections. 
A consequence of this is that
the slow roll parameters and the number of efolds will depend on
the beta-functions, and differ from the respective tree-level
expressions. However, when calculating the spectral index and
tensor-to-scalar ratio, all corrections exactly cancel.  This can be
shown, expanding in the small parameter defying the inflationary regime, without explicitly solving the RGEs. This is our main result,
the inflationary predictions for $\alpha$ and $\xi$ attractors are not
affected by quantum corrections (to leading order in the $1/N$
expansion). This generalize our previous work on Higgs inflation \cite{Fumagalli:2016lls}
to the larger class of Cosmological Attractors. At the same time it extends beyond the tree level the unified picture provided in \cite{Galante:2014ifa}.

There is one caveat which allows quantum
corrections to become important. It may be that the potential
develops an extremum because of the running; this is purely a quantum
effect since the tree level potential has no extremum in the
inflationary regime.  This coincides with the breakdown of our
perturbative expansion and analytical control is lost. This allows for hilltop or inflection point
inflation which are sensitive to the loop corrections. However, these cases are realized for very fine-tuned values of the parameters and they can be studied numerically for specific inflationary scenarios. 

The conclusions remain valid as long as the kinetic term
and the potential can be written in the form \eqref{eq:kin and pot}
and the leading log in the effective potential is minimized by an RG
function satisfying \eqref{eq:dt/drho}.
This turned out to be correct provided the theory is perturbatively renormalizable, in both the low/high field regime; which is the \textit{sine qua non} for our full discussion. 
\section*{Acknowledgements}

I am really grateful to Marieke Postma for enlightening discussions, valuable input and useful corrections to the manuscript. I am sincerely thankful to Mario Galante for a careful reading of an early draft of this note. I would also thank Tomislav Prokopec, Cliff Burgess, Jose Espinosa, Sander Mooij, Robbert Rietkerk and Sergio Tapias  for fruitful conversations.  
The authors is funded by the Netherlands Foundation for Fundamental Research of Matter (FOM) and the Netherlands Organisation for Scientific Research (NWO).

\bibliographystyle{utphys}
\bibliography{biblio2}

\end{document}